\documentclass[pre,aps,twocolumn,superscriptaddress,amsmath,showpacs,floatfix]{revtex4}
\usepackage{graphics,epsfig}
\usepackage{amssymb}

\begin{document}
\title{``Water-free'' computer model for fluid bilayer membranes}

\author{Oded Farago}
\email{farago@mrl.ucsb.edu}
\affiliation{Materials Research Laboratory, University of
  California, Santa Barbara, CA 93106}
\altaffiliation{Also at Department of Physics, Korea Advanced Institute of
Science and Technology (KAIST), 373-1 Kusong-dong, Yusong-gu, Taejon
305-701, South Korea.}

\begin{abstract}
\vspace{0.5cm}  
  We use a simple and efficient computer model to investigate the
  physical properties of bilayer membranes. The amphiphilic molecules
  are modeled as short rigid trimers with finite range pair
  interactions between them. The pair potentials have been designed to
  mimic the hydrophobic interactions, and to allow the simulation of
  the membranes without the embedding solvent as if the membrane is in
  vacuum. We find that upon decreasing the area density of the
  molecules the membrane undergoes a solid-fluid phase transition,
  where in the fluid phase the molecules can diffuse within the
  membrane plane. The surface tension and the bending modulus of the
  fluid membranes are extracted from the analysis of the spectrum of
  thermal undulations. At low area densities we observe the formation
  of pores in the membrane through which molecules can diffuse from
  one layer to the other. The appearance of the pores is explained
  using a simple model relating it to the area dependence of the free
  energy.

\end{abstract}

\maketitle

\section{Introduction}

When amphiphilic molecules such as lipids are brought into contact
with water they tend to arrange so as to shield their ``oily''
hydrocarbon tail from the aqueous environment while exposing their
hydrophilic head to the water. One of the simplest structures formed
in this way is of a bilayer membrane - a double sheet of surfactants
separating two aqueous phases \cite{israelachvili}. Bilayer membranes
are common in biological systems \cite{alberts}. Living cells are
separated from their extra-cellular surroundings by plasma membranes
that control the transport of material into and out of the cell
\cite{sackmann,bloom}. Most biological membranes are found in the
fluid phase where the lipids comprising the bilayer can diffuse freely
in the membrane plane.  Another characteristic feature of lipid
bilayers is their high flexibility which allows for large
thermally-excited undulations \cite{safran,lipowsky_sackmann}.  The
fluidity and low rigidity of membranes are important for many of their
biological properties, such as their ability to change their shape
easily and the possibility of proteins to insert themselves into the
membrane \cite{gennis}.

The thickness of membranes is comparable to the size of the
constituting surfactant molecules (typically on the nanometer scale),
while their lateral extension can greatly exceed their thickness and
reach up to several micrometers. Consequently, coarse-grain
phenomenological models, such as Ginzburg-Landau free energy
functionals \cite{gompper_schick} or the effective surface Hamiltonian
\cite{safran,lipowsky_sackmann,gompper_schick,helfrich}, have been
used in order to study the physical properties of membranes, as well
as of other interfaces (like surfactants monolayer in microemulsions
or vapor-fluid interfaces). In those theories the bilayer membrane is
treated as a smooth continuous surface, and its elastic energy is
related to the membrane area and the local curvatures. These theories
have been very successful in describing the shape and phase diagrams
of bilayer membranes \cite{seifert_lipowsky,wortis}.

Phenomenological models describe the mesoscopic physical behavior of
interfaces and membranes, but do not allow one to approach these
systems on the molecular level. Many theories have been developed in
an attempt to understand how the mesoscopic behavior emerges from the
microscopic entities and the interactions between them. These theories
include lattice ``Ising-like'' models \cite{gompper_schick}, molecular
theories of the hydrocarbon chain packing \cite{szleifer}, theories
including the effect of electrostatic interactions \cite{andelman},
and density functional theories \cite{chacon}. The most microscopic
detailed approach is employed in some computer simulations where the
amphiphiles and water, and the interactions between them are modeled
explicitly in full detail \cite{atomistic}.  Since these simulations
require an enormously large computing time, they are restricted to
fairly small systems consisting of 50-200 amphiphiles, and can be
utilized to investigate phenomena occurring on short time scales of a
few nanoseconds. In order to study mesoscale phenomena it is therefore
necessary to dispense with some of the microscopic details in the
simulations and use simplified models \cite{simplified_review}. A
number of such simplified computer models have been devised by several
groups. In these models the structure of the surfactant molecules is
represented in a ``coarse-grained'' manner where a number of atoms are
grouped together into a single site. The first level of
coarse-graining is obtained by replacing the water molecules and the
$\rm{CH_2}$ groups of the hydrocarbon chain by unified atoms
\cite{simplified1a,simplified1b,simplified1c}. This can reduce the
number of atoms per lipids to about 50.  Much more simplified models,
in which the amphiphiles consists of only 5-10 atoms, were also
presented \cite{smit,goetz,shelley}. In these latter models the
electrostatic potentials are usually ignored and the potentials of the
chemical bonds are greatly simplified.  At this level of
simplification it is obviously impossible to address specific lipids
systems, but rather the more general properties of self-assembling
systems.

The size of amphiphilic systems which can be simulated using
simplified models is constantly growing by virtue of the availability
of inexpensive and powerful commodity PC hardware and due to the
development of new simulation techniques such as Dissipative Particle
Dynamics (DPD) \cite{dpd}. Simulations of model systems consisting of
$N\gtrsim 1000$ lipids have been recently reported in the literature
\cite{dpd,lopez,muller1}. The major restriction on the size of the
systems in these simulation stems from the large number of atoms
included in the simulation cell which is typically an order of
magnitude larger than the number of amphiphilic molecules. The low
ratio between the number of lipids and the {\em total}\/ number of
atoms is due to two factors. The first one is the number of atoms
comprising each lipid molecule which, as discussed above, can vary
from 50 to 5 depending on the level of simplification employed in the
simulations. The second factor is the number of water molecules in the
simulation cell. In bilayers simulations the typical number ratio of
water to lipid molecules is in the range of 10 to 30
\cite{atomistic,simplified1a,simplified1b,simplified1c,smit,goetz,shelley}.
A great fraction of the computing time is, thus, ``wasted'' on the
simulations of the water even when the water molecules are represented
by a single (unified) atom. Only very few models have so far been
proposed in which the amphiphiles are simulated without the presence
of water. The major difficulty in establishing such ``water-free''
models is the need to mimic the hydrophobic effect that prevents the
amphiphilic molecules from leaving the aggregate into the solvent.
Drouffe at al.~\cite{drouffe} and Noguchi at al.~\cite{noguchi} have
used {\em ad hoc}\/ multibody potentials to overcome this problem.
With the aid of these nonphysical potentials they have managed to
observe the formation of fluid vesicles in their simulations. La Penna
at al.~\cite{lapenna} have studied a water-free flat bilayer model
with Lennard-Jones (LJ) potentials that depend on the relative
orientation of the lipids (and which are closely related to the
Gay-Berne \cite{gay_berne} potentials used in liquid crystals
simulations). With this model they have been able to simulate bilayer
membranes in both the solid and the gel phases. Fluid membranes,
however, were found unstable against lipids evaporation from the
membrane plane.

In this paper we present an exceptionally simple computer model of a
{\em fluid}\/ bilayer membrane. Our model has the following features:
(a) It is a water-free model, i.e., we simulate the membrane without
the presence of water. (b) The ``lipids'' forming the membranes
consist of only three atoms, one representing the hydrophilic
headgroup and the other two the hydrophobic tail. These three atoms
are ``glued'' to each other to form a rigid linear trimer (the lipid),
and have no additional interactions between them. (c) The different
lipids interact through finite range (truncated) LJ interactions
between their three sites. The parameters of the LJ potentials are
fixed and do {\em not}\/ depend neither on the relative orientation of
the lipids (as in Ref.~\cite{lapenna}), nor on their local density
(i.e., there are no multibody interactions in our model). The above
mentioned properties make our membrane model computationally very
efficient (albeit a less ``flexible'' one in comparison to other
simplified models with more interaction sites per amphiphile). To
investigate the statistical mechanical properties of the membrane, we
have performed a set of Monte Carlo (MC) simulations where for each MC
run we have fixed the temperature, the number of lipids, and the
projected area of the membrane. The projected area serves as the
control parameter in our simulations, and we have investigated the
phase behavior of the membrane as a function of it.  We found that
upon increasing the projected area (i.e., reducing the area density of
the lipids) the membrane undergoes a solid-fluid phase transition. In
the solid phase the lipids are not mobile and they pack in a hexagonal
order. In the fluid phase the lipids are free to diffuse in the
membrane plane. We have measured the spectrum of thermal undulations
of the fluid membranes from which we have extracted the surface
tension and the bending modulus that characterize the elastic behavior
of the membrane. At low area densities we found another transition
from negative to positive surface tension, accompanied by the
formation of pores in the membranes. Such a behavior is indeed
predicted by theoretical arguments \cite{litster,netz,muller2}

The paper is organized as follows: In section \ref{model} we present
our computer model, and discuss the details of the simulations. In
section \ref{simulations} we describe the physical properties of the
systems as obtained by the simulations. The section is divided into
three subsections dealing, respectively, with the phase diagram of the
system, its spectrum of thermal undulations and elastic properties,
and the appearance of holes in membranes with large projected area. We
summarize and discuss the results in section \ref{summary}

\section{Details of the Model and the Simulations}
\label{model}

\begin{figure}
\epsfig{file=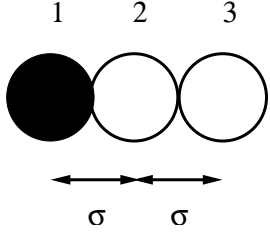,width=3.5cm}
\caption{A schematic picture of a lipid molecule in our model system - a 
  rigid linear trimer consisting of three atoms whose centers are
  separated a distance $\sigma$ apart. The atom labeled 1 (solid
  circle) represents the hydrophilic head of the lipid, while the
  atoms labeled 2 and 3 (open circles) represent the hydrophobic
  tail.}
\label{lipid}
\end{figure}

\begin{figure}
\epsfig{file=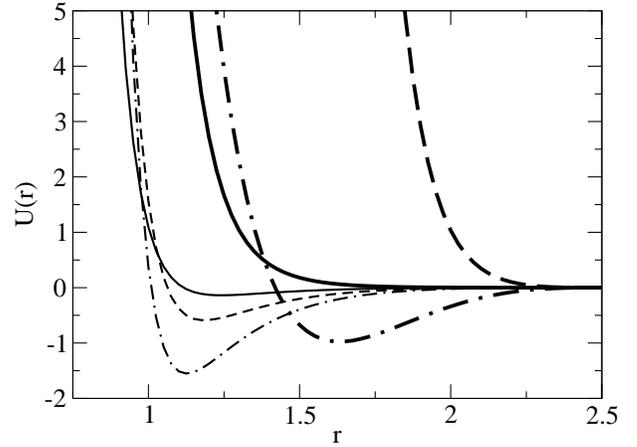,width=8cm}
\caption{The different pair potentials $U_{ij}(r)$ used in our model: 
  $U_{11}$ (solid line), $U_{22}$ (dashed line), $U_{33}$
  (dotted-dashed line), $U_{12}$ (bold solid line), $U_{13}$ (bold
  dashed line), and $U_{23}$ (bold dotted-dashed line). The distance
  $r$ is in units of $\sigma$ (see definition in text), while the
  potentials $U_{ij}$ are in $kT$ units. The parameters $\sigma_{ij}$
  and $\epsilon_{ij}$ [see Eqs.({\protect \ref{LJ12}})--({\protect
    \ref{fullpotentials}})] are as follows: $\sigma_{11}=1.1\sigma$,
  $\sigma_{22}=1.05\sigma$, $\sigma_{33}=\sigma$,
  $\sigma_{12}=1.15\sigma$, $\sigma_{13}=1.4\sigma$,
  $\sigma_{23}=0.525\sigma$, $\epsilon_{11}=0.1875kT$,
  $\epsilon_{22}=1.75kT$, $\epsilon_{33}=1.875kT$,
  $\epsilon_{12}=1.1375kT$, $\epsilon_{13}=200kT$, and
  $\epsilon_{23}=375kT$.}
\label{potentials}
\end{figure}

The lipids in our model system consist of three spherical atoms
connected to form a linear trimer. The lipid molecules are rigid -
they do not bend and the distance $\sigma$ between the center of the
atoms is {\em fixed}\/ (see fig.~\ref{lipid}). We set $\sigma=1$ as
our unit length scale throughout this paper. We shall label the three
atoms forming each lipid as 1, 2, and 3. Atom 1 represents the
hydrophilic head of the lipid, while atoms 2 and 3 represent its
hydrophobic tail. The different lipids interact with each other via
spherically symmetric pair potentials between their constituting
atoms. The pair potential $U_{ij}(r)$ depicts the interactions between
atom $i$ and atom $j$ of two different molecules separated a distance
$r$ apart. The pair potentials $U_{12}$ and $U_{13}$ describe the
interaction between hydrophobic and hydrophilic particles. They are
given by the purely repulsive LJ potential
\begin{equation}
U_{12}^{\rm LJ}(r)=4\epsilon_{12}\left(\frac{\sigma_{12}}{r}\right)^{12},
\label{LJ12}
\end{equation}
and
\begin{equation}
U_{13}^{\rm LJ}(r)=4\epsilon_{13}\left(\frac{\sigma_{13}}{r}\right)^{18}.
\label{LJ13}
\end{equation}
The pair potentials $U_{11}$, $U_{22}$, and $U_{33}$ describe the
interactions between two similar atoms, both either hydrophobic or
hydrophilic. They are given by the attractive LJ potentials
\begin{equation}
U_{ii}^{\rm LJ}(r)=
4\epsilon_{ii}\left[\left(\frac{\sigma_{ii}}{r}\right)^{12}
-\left(\frac{\sigma_{ii}}{r}\right)^{6}\right],
\label{LJii}
\end{equation}
where $i=1,2,3$. Finally, the interaction between the hydrophobic
particles 2 and 3 is also depicted by an attractive LJ potential, but of 
the form
\begin{equation}
U_{23}^{\rm LJ}(r)=
4\epsilon_{23}\left[\left(\frac{\sigma_{23}}{r}\right)^{2}
-\left(\frac{\sigma_{23}}{r}\right)\right].
\label{LJ23}
\end{equation}
All pair potentials are truncated at the same cut-off $r_c=2.5\sigma$,
and the discontinuity at $r_c$ is avoided by adding extra terms to the
LJ potentials that ensure the vanishing of the potential, as well as
of its first and second derivative, at $r=r_c$. The final form of the
pair potentials is thus given by
\begin{eqnarray}
U_{ij}(r)=U_{ij}^{\rm LJ}(r)&-&U_{ij}^{\rm LJ}(r_c)-
\frac{\partial U_{ij}^{\rm LJ}(r)}{\partial r}\Biggm|_{r=r_c}(r-r_c)
\nonumber \\ 
\ &-&\frac{1}{2}\frac{\partial^2 U_{ij}^{\rm LJ}(r)}
{\partial r^2}\Biggm|_{r=r_c}(r-r_c)^2.
\label{fullpotentials}
\end{eqnarray}
The different pair potentials are depicted in fig.~\ref{potentials}.
The values of the parameters $\sigma_{ij}$ (in units of $\sigma$) and
$\epsilon_{ij}$ (in $kT$ units, where $T$ is the temperature and $k$
the Boltzmann constant) used in the simulations are summarized in the
caption on fig.~\ref{potentials}.

The pair potentials in our computer model have been designed to allow,
on the one hand, the diffusion of molecules in the plane of the
membrane but to restrict, on the other hand, their motion in the third
direction.  We have tested various models before we arrived to the one
that we have used in the simulations. The original idea was to use
dimers with one hydrophilic and one hydrophobic particles, and to
describe the interactions between them by 6-12 LJ potentials
[Eq.(\ref{LJii})] and a 12-power repulsive potential
[Eq.(\ref{LJ12})], depending on whether the atoms are of the same or
different species. It turned out that the membranes depicted by such a
model were unstable against the extraction of molecules from the
membrane plane. To increase the membrane stability we added a third
hydrophobic atom to the lipids. The pair interactions between this
atom and the other two atoms are described by different forms of LJ
potentials: For the interaction with the hydrophilic atom labeled 1 we
use the more repulsive 18-power LJ potential $U_{13}$
[Eq.(\ref{LJ13})], while for the interaction with the hydrophobic atom
labeled 2 we use the 1-2 LJ potential $U_{23}$ [Eq.(\ref{LJ23})]. The
former potential establishes a strong repulsion between the
hydrophobic and the hydrophilic parts of the lipids, thus reducing
significantly (eliminating on the time scale of the simulations) the
escape probability of molecules (more on this point in the next
paragraph).  The latter has a very shallow minimum which allows a
greater mobility of the lipids in the membrane plane (by making small
the energy changes due to a relative motion of the lipids with respect
to each other). We have gone through a rather lengthy ``trial and
error'' process of fine tuning the parameters $\sigma_{ij}$ and
$\epsilon_{ij}$ which control the range of pair repulsion and the
depth of the attractive potential wells. Their values have been set to
(a) make favorable the alignment of molecules next to each other at a
distance slightly larger than $\sigma$, and (b) to make the attraction
between molecules sufficiently strong to support the stability of the
membrane, but not too strong to the extent that would entirely prevent
the diffusion of the lipids.

It is not an easy task to form a fluid bilayer sheet in a model system
that does not contain water. Membranes become fluid at low area
densities and high temperatures, and under these conditions the lipids
tend to escape quite easily from the membrane plane. It is the water
that confines the lipids to the membrane. In the absence of water
molecules this role has to be played by the hydrophobic heads which
must form some sort of geometric or dynamic constraint for the
extraction of lipids. In our model we establish such a constraint by
making the excluded volume part of the pair potentials $U_{ij}$ {\em
  non-additive}\/, namely we make the size of a particle $i$ ``seen''
by another particle $i$ smaller than its size as seen by a particle
$j$ of a different species. We can define the distance $a_{ij}$ at
which the pair potential between them $U_{ij}=kT$ as a measure for the
range of hard core repulsion between the two particles $i$ and $j$.
(It is unlikely to find a pair $i$ and $j$ separated by a smaller
distance.) It is customary to regard $a_{ii}$ as the diameter of atom
$i$ and, with this interpretation, to expect for the additivity of the
hard core diameters, i.e., to have $a_{ij}\simeq (a_{ii}+a_{jj})/2$
for $i\neq j$.  In our model we do not find this property (see
fig.~\ref{potentials}).  The pair potentials in our system describe
the effective interactions between the different atoms and they
include the effect of the water molecules which are not simulated
explicitly. Therefore, there is no a priory reason why the effective
diameters associated with different particles should be strictly
additive. The increased range of hard core repulsion between the
hydrophilic atom 1 and the hydrophobic atoms 2 and 3 is designed to
compensate for the absence of water from the simulation cell.

The simulations were performed with membranes consisting of $N=1000$
lipids (500 lipids in each layer) with periodic boundary conditions in
the membrane $(x,y)$ plane, and with no boundaries in the normal
$z$-direction. Subsequent MC configurations were generated by two
types of move attempts: translations of lipids and rotations around
the mid (second) atom. The MC unit time is defined as the time
(measured in number of MC configurations) in which, on the average, we
attempt to move and rotate each molecule once. The acceptance
probability of both types of moves was approximately half. We
performed a set of simulations of membranes with the same temperature
$T$ and number of lipids $N$, and with varying projected areas. For
each value of the projected area we studied 8 different membranes
starting at different initial configurations. The initial
configurations were created by randomly placing 500 lipids in two
layers with a vertical (along the normal $z$ direction) separation
$\sigma$ between the atoms labeled 3 in the two layers, and with all
the lipids oriented normal to the membrane plane.  The initial
configurations were then ``thermalized'' over a period of $5\cdot10^5$
MC time units, followed by a longer period of $6\cdot10^6$ time units
during which quantities of interest were evaluated. The duration of
the MC runs is substantially larger than the relaxation time which we
estimated in various ways: As a first approximation for the relaxation
time we used the time it took the potential energy of the membrane to
saturate from its high initial value (resulting from overlap of
particles in the random initial configuration) to a final ``typical''
value. This time was of the order of $10^4$ MC time units. An
independent estimate of the relaxation time was obtained from a study
of the spectrum of thermal undulations of the membranes (see more
details, later in the text).  Inspection of the autocorrelation
function of the amplitude of the longest wavelength mode led to a
similar estimate of $10^4$ time units for the relaxation time. A more
conservative estimate can be obtained from measurements of the
self-diffusion constant of the lipids in the fluid phase (see, again,
later in the text). The relaxation time can be associated with the
time it takes a molecule to diffuse a distance equal to the pair
potentials cut-off ($2.5\sigma$).  The relaxation time obtained using
this criterion was an order of magnitude larger ($\sim 10^5$ MC time
units), still smaller than the equilibration time, and much smaller
than the total length of the simulations.

\section{Simulation Results}
\label{simulations}

\subsection{Phase Diagram}

The projected area of the membranes in the simulations ranges from
$A_p=L_p^2=(26.875)^2$ to $A_p=(30.625)^2$ with intervals of $\Delta
L_p=0.625$. For all area densities we measured the self-diffusion
constant of the molecules relative to the diffusion of the center of
mass, defined by
\begin{widetext}
\begin{equation}
D\equiv\lim_{t\rightarrow\infty}\frac{\Delta {r'(t)}^2}{4t}
\equiv\lim_{t\rightarrow\infty}\frac{1}{4Nt} \sum_{i=1}^{N}\left[
\left(\vec{r}_i(t)-\vec{r}_{\rm CM}(t)\right)-
\left(\vec{r}_i(0)-\vec{r}_{\rm CM}(0)\right)\right]^2,
\label{diffusion} 
\end{equation}
where $\vec{r}_i(t)$ denotes the
position of the $i$-th lipid (defined by the position of its mid atom)
at time $t$, while $\vec{r}_{\rm
  CM}(t)$ denotes the position of the center of mass of lipids 
\cite{remark0}. We
have also measured the self-diffusion coefficient {\em in the membrane
  plane}\/, defined by: 
\begin{equation}
D_{x-y}\equiv\lim_{t\rightarrow\infty}\frac{1}{4Nt}
\sum_{i=1}^{N}\left\{\left[ \left(x_i(t)-x_{\rm CM}(t)\right)-
\left(x_i(0)-x_{\rm CM}(0)\right)\right]^2+
\left[ \left(y_i(t)-y_{\rm CM}(t)\right)- \left(y_i(0)-y_{\rm
CM}(0)\right)\right]^2\right\}, 
\label{diffusionxy} 
\end{equation} 
where $x$ and $y$ denote Cartesian coordinates. (In all the
simulations the membranes lied in the $(x,y)$ plane, while fluctuating
in the normal $z$-direction.) As the lipids can only diffuse within
the plane of the membrane, we found no difference between $D$ and
$D_{x-y}$.
\end{widetext}

\begin{figure}[h]
\epsfig{file=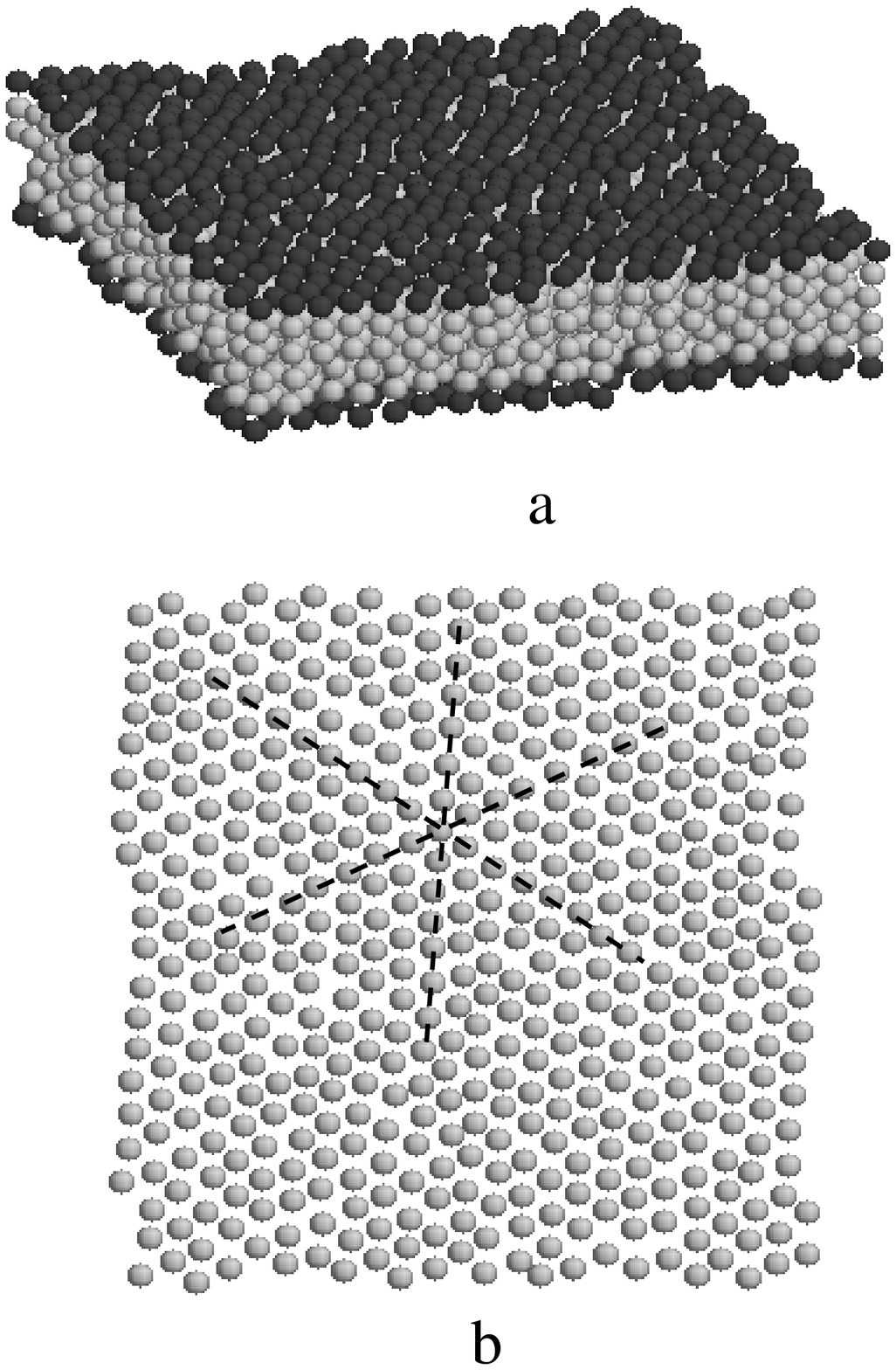,width=7cm}
\caption{(a) Equilibrium configuration of a solid membrane with 
  $A_p=(26.875)^2$. The atoms labeled 1 (the ``hydrophilic'' atoms)
  are shown as black spheres of diameter $\sigma$, while the grey
  shaded spheres depict the (hydrophobic) atoms labeled 2 and 3, and
  are of diameter $\sigma$ as well. (b) A top view of the plane of mid
  (labeled 2) atoms of the membrane upper layer.}
\label{solid}
\end{figure}

At low projected area (high area density) we found the membrane in a
solid phase characterized by two features: (a) The diffusion constant
of the lipids is vanishingly small. [The root mean square displacement
$\sqrt{\Delta r'^2}$ has barely changed during the course of the
simulations, and it has never exceeded the typical distance between
neighbor molecules ($\sim \sigma$).] (b) The lipids arrange themselves
in a hexagonal order in the membrane plane. A typical equilibrium
configuration of a membrane with $A_p=(26.875)^2$ is shown in
fig.~\ref{solid} (a). A top view of the plane of mid (labeled 2) atoms
of the membrane upper layer, revealing the hexagonal order of the
lipids, is shown in fig.~\ref{solid} (b).  The lattice imperfections
observed at fig.~\ref{solid} (b) should be mainly attributed to the
incommensurability of the 500 sites hexagonal lattice with the square
simulation cell.

\begin{figure}[t]
\epsfig{file=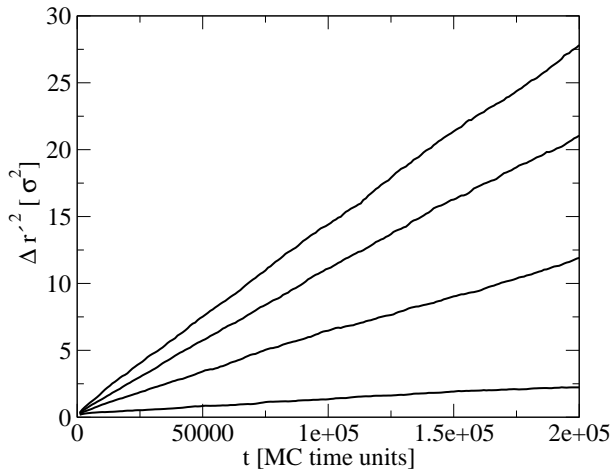,width=8cm}
\caption{The lipids mean square displacement $\Delta {r'}^2$ (measured in 
  $\sigma ^2$ units) as a function of the time (measured in MC time
  units) for fluid membranes with (from bottom to top)
  $A_p=(28.125)^2$, $A_p=(28.75)^2$, $A_p=(29.375)^2$, and
  $A_p=(30.0)^2$.}
\label{diffusioncurves}
\end{figure}

\begin{figure}
\epsfig{file=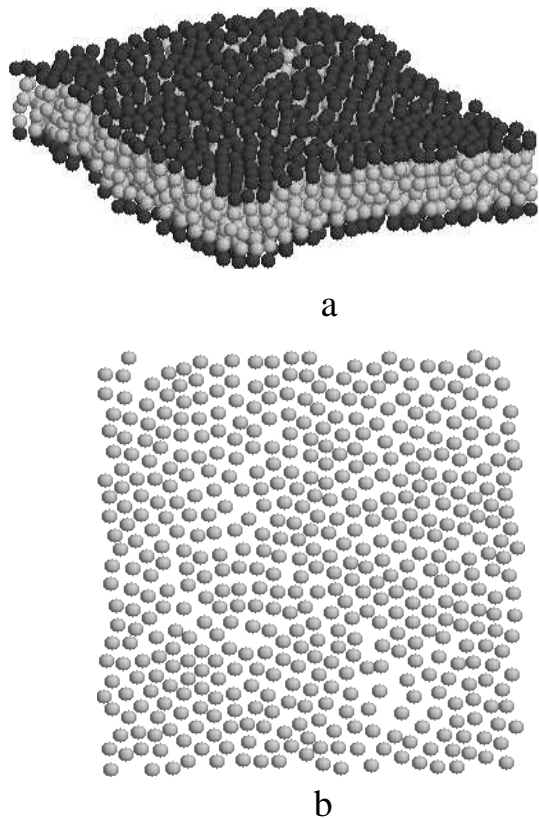,width=7cm}
\caption{(a) Equilibrium configuration of a fluid membrane with 
  $A_p=(28.75)^2$. Black and grey atoms (of diameter $\sigma$) depict
  hydrophilic (labeled 1) and hydrophobic (labeled 2 and 3) atoms,
  respectively. (b) A top view of the plane of mid (labeled 2) atoms
  of the membrane upper layer.}
\label{fluid}
\end{figure}

At larger values of the projected area [$A_p\geq(28.125)^2$] we found
the membranes in a fluid phase. The main feature that distinguishes
fluid from solid membranes is the diffusion of the lipids. In
fig.~\ref{diffusioncurves} we plot lipids mean square displacement
$\Delta {r'}^2$ [see definition in Eq.(\ref{diffusion})] as a function
of the simulation time $t$ for fluid membranes with different area
densities.  The slope of the asymptotically linear curves is four
times larger than the self-diffusion constant $D$. One can observe the
growth of $D$ with the increase of the projected area - a rather
expected observation as the increase of the projected area means more
room for the molecules to move. A typical equilibrium configuration of
a fluid membrane with $A_p=(28.75)^2$ is depicted in fig.~\ref{fluid}
(a).  Another characteristic feature of the fluid membranes is the
loss of in-plane hexagonal order, as demonstrated in fig.~\ref{fluid}
(b) [compare with fig.~\ref{solid} (b)].

\begin{figure}[t]
\epsfig{file=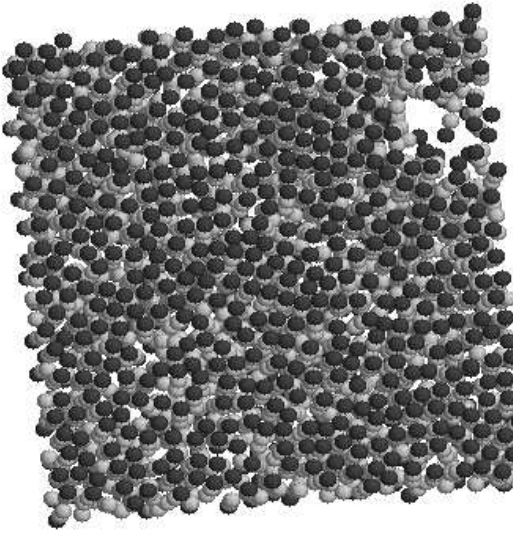,width=7cm}
\caption{Equilibrium configuration of a fluid membrane with 
  $A_p=(30.0)^2$ having a pore on its upper right corner.}  
\label{pore}
\end{figure}

\begin{figure}
\epsfig{file=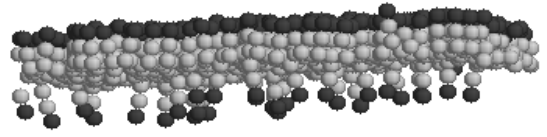,width=7cm}
\caption{Another view at the membrane depicted in 
  fig.~{\protect \ref{pore}}. Here we show only half of the lipids
  which were originally located on the upper leaflet of the bilayer.}
\label{flipflops}
\end{figure}

The membranes with $A_p=(30.0)^2$ exhibited an interesting feature -
they developed pores, as demonstrated in the configuration shown in
fig.~\ref{pore}. These pores tended to appear irregularly in the
membrane with a characteristic time scale $\tau\gtrsim 2\cdot10^5$ for
the formation of a pore, and a typical pore life time of a few
thousand time units. Another interesting phenomenon which we observed
for this value of $A_p$ and did not observe at lower projected areas
was the occurrence of ``flip-flops'' - the transition of lipids from
one layer to the other. In fig.~\ref{flipflops} we look at the same
membrane depicted in fig.~\ref{pore}. In this figure, however, we plot
only the 500 lipids that were located in the upper layer in the
initial configuration.  About 30 of them have managed to diffuse from
the upper to the lower layer during the course of the simulations. A
similar (although not necessarily identical) number of lipids have
moved in the opposite direction. Trans-bilayer diffusion is an
important process in real bilayer membranes \cite{sackmann2}. To allow
for uniform bilayer growth, some of the lipids must be transfered from
one leaflet to the other during the self-assembly process. When a flat
bilayer is bent to form a spherical vesicle, the area of the inner
layer becomes smaller than the area of the outer layer, and it is the
transition of lipids from the former to the latter that balances their
area densities. It has been suggested, based on experiments
\cite{matsuzaki} and computer simulations \cite{okazaki}, that that
the formation of pores and the flip-flop motion are closely
interconnected. According to these studies, the pores provide a
transverse diffusion conduit for the lipids, through which their
hydrophilic headgroups cross the hydrophobic region of the membrane.
Our study supports the conjecture about the possible relation between
pores and flip-flops. In section \ref{poresection} we discuss the
origin of the formation of the pores. We show that it is associated
with the change in the sign of the membrane surface tension - from a
negative to a positive value. Once the surface tension attains a
sufficiently large positive value, the energy involved with the
formation of the pores is compensated by the reduction in elastic
energy.

While the formation of the pores allowed the diffusion of molecules
between the two layers, we did not observe [in membranes with
$A_p=(30.0)^2$] that pores also lead to the extraction of molecules
from the membrane. It is, however, possible that the disassociation of
pore-forming membranes occurs on time scales larger than the duration
of our simulation. Fast disintegration of the membrane was observed
when the projected area was increased to $A_p=(30.625)^2$, which was,
therefore, the largest projected area set for the membranes in our
study.

\subsection{Elasticity and Thermal Undulations}

On length scales larger than the membrane thickness, the bilayer can
be modeled as a smooth continuous sheet. The thermal undulations of
the bilayer can be studied with Helfrich Hamiltonian \cite{helfrich}
relating the elastic energy to the shape of the membrane:
\begin{equation} 
{\cal H}=\int_{S} dA\,
\left[\gamma+\frac{1}{2}\kappa
\left(c_{1}+c_{2}-2c_{0}\right)^2+\kappa_{G}c_{1}c_{2}\right].
\label{helfhamiltonian} 
\end{equation} 
The integration in the above equation is carried over the whole
surface of the membrane. Three elastic moduli are involved with the
Helfrich Hamiltonian: the surface tension $\gamma$, the bending
modulus $\kappa$, and the saddle-splay modulus $\kappa_G$. The
quantities $c_1$ and $c_2$ appearing in the above equation are the
local principle curvatures of the surface (see a rigorous definition
in Ref.\cite{fdavid}) which are surface invariants with respect to
similarity transformations (translations and rotations), while $c_0$
is the spontaneous curvature of the surface. For flat bilayers
$c_0=0$. It is customary to dispense with the use of the local
curvatures in favor of two other (local) invariants: the mean
curvature $H\equiv (c_1+c_2)/2$, and the Gaussian curvature $K\equiv
c_1c_2$.  If one only considers fluctuations which do not change the
topology of the membrane, then the total energy associated with the
last term in Eq.(\ref{helfhamiltonian}) is a constant\cite{helfrich}.
We, thus, arrive to the following more simplified form of
Eq.(\ref{helfhamiltonian}): \begin{equation} {\cal H}=\int_{S} dA\,
  \left(\gamma+2\kappa H^2\right). \label{helfhamiltonian2}
\end{equation}

There are various ways to parameterize the surface. One of them is the
Monge representation, where the surface is represented by a height
function, $z=h(x,y)$, above a reference $x-y$ plane. For a nearly flat
surface, i.e., when the derivatives of the height function with
respect to $x$ and $y$ are small -- $h_x, h_y\ll 1$, one obtains the
following approximation for Eq.(\ref{helfhamiltonian2}):
\begin{equation}
{\cal H}=\int_{} dxdy\, \left[\frac{1}{2}\gamma
\left(h_x^2+h_y^2\right)+\frac{1}{2}\kappa
\left(\nabla^2 h\right)^2\right].
\label{helfhamiltonian3}
\end{equation}
Note that unlike Eq.(\ref{helfhamiltonian2}), the integral in
Eq.(\ref{helfhamiltonian3}) runs over the reference $(x,y)$ surface
rather than over the actual surface of the membrane.

Equations (\ref{helfhamiltonian})--(\ref{helfhamiltonian3}) are
expected to be valid only on length scales larger than the thickness
of the membrane. The undulatory motion on smaller length scales (which
we did not investigate in this study) is dominated by the so called
``protrusion modes'' \cite{protrusion}. In our simulations the profile
of the bilayers was defined by mapping the system with linear size (of
the projected area) $L$ onto an $8\times 8$ grid whose mesh size
$l=L/8$ is indeed larger than the typical width of the membrane. The
local height of the bilayer was then defined as the average of the
local heights of the two layers. The latter were evaluated by the mean
height of the lipids (whose positions were identified with the
coordinates of their mid atoms) belonging to each layer, which were
instantaneously located inside the local grid cell. The discretized
form of Hamiltonian (\ref{helfhamiltonian3}) is
\begin{equation}
{\cal H}=a_0\sum_{\vec{r}} \left[\frac{1}{2}\gamma
\left(h_x^2+h_y^2\right)+\frac{1}{2}\kappa
\left(\nabla^2 h\right)^2\right],
\label{helfhamiltonian4}
\end{equation}
where summation goes over the discrete grid coordinates, and $a_0=l^2$
is the area of the grid cells. In Fourier coordinates we define
\begin{equation}
h\left(\vec{r}\right)=\frac{l}{L}\sum_{\vec{q}}h_{\vec{q}}\,
e^{i\vec{q}\cdot\vec{r}},
\end{equation}
and 
\begin{equation}
h_{\vec{q}}=\frac{l}{L}\sum_{\vec{r}}h\left(\vec{r}\right)
e^{-i\vec{q}\cdot\vec{r}},
\label{fourier}
\end{equation}
where the two-dimensional wavevector $\vec{q}$ has $8^2=64$ discrete
values satisfying $\{q_x,q_y=2\pi n/L$, $n=-4,-3,\ldots,2,3\}$. In
Fourier space the different modes decouple:
\begin{equation}
{\cal H}=\frac{a_0}{2}\sum_{\vec{q}} \left[\gamma |\vec{q}\,|^2+
\kappa|\vec{q}\,|^4\right]|h_{\vec{q}}\,|^2,
\label{helfhamiltonian5}
\end{equation}
and, by invoking the equipartition theorem, we find that the mean
square amplitude of the mode $\vec{q}$ (the ``spectral intensity'')
\begin{equation}
a_0\langle |h_{\vec{q}}\,|^2\rangle=\frac{kT}{\left
(\gamma |\vec{q}\,|^2+\kappa|\vec{q}\,|^4\right)}.
\label{qamplitude} 
\end{equation}

\begin{figure}
\epsfig{file=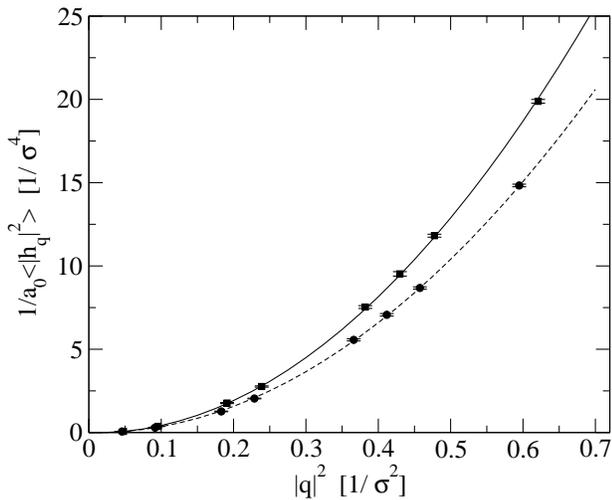,width=8cm}
\caption{The inverse of the spectral intensity for undulatory modes
  $1/a_0\langle|h_q|^2\rangle$ as a function of the square wave number
  $q^2$ for membranes with $A_p=(28.75)^2$ (squares) and
  $A_p=(29.375)^2$ (circles).}  
\label{spectrum} 
\end{figure}

The instantaneous amplitudes of the different $q$-modes were evaluated
using Eq.(\ref{fourier}) once in every 100 MC time units, and were
averaged over the course of the simulations. To extract the values of
$\gamma$ and $\kappa$ we used the inverse from of
Eq.(\ref{qamplitude})
\begin{equation}
\frac{1}{a_0\langle|h_{\vec{q}}\,|^2\rangle}=\frac{\left
(\gamma |\vec{q}\,|^2+\kappa|\vec{q}\,|^4\right)}{kT},
\label{qamplitude2} 
\end{equation}
and plotted $1/a_0\langle|h_{\vec{q}}\,|^2\rangle$ and a function of
$|\vec{q}\,|^2$. The results of the spectral analysis of the
undulations for fluid membranes with $(A_p=28.75)^2$ (squares) and
$A_p=(29.375)^2$ (circles) are presented in fig.~\ref{spectrum}. The
error bars represent one standard deviation in the estimates of the
averages. The curves depict the best fit of the numerical data to
Eq.(\ref{qamplitude2}), obtained when $\gamma$ and $\kappa$ take the
following values:
\begin{eqnarray}
\gamma&=&-1.4\pm 0.2\ \frac{kT}{\sigma^2}
\nonumber\\
\kappa&=&54\pm 2\ kT,
\label{moduli28.75}
\end{eqnarray}
for $A_p=(28.75)^2$, and,
\begin{eqnarray}
\gamma&=&-0.6\pm 0.2\ \frac{kT}{\sigma^2} 
\nonumber\\
\kappa&=&42\pm 2\ kT,
\label{moduli29.375}
\end{eqnarray}
for $A_p=(29.375)^2$. We verified the validity of
Eq.(\ref{qamplitude2}) by attempting to fit our data to other
polynomial functional forms, including a constants and a
$|\vec{q}|\,^6$ terms. The contributions of these terms to the fit
were small, and did not result a significant change in our estimates
of $\gamma$ and $\kappa$, based on Eq.(\ref{qamplitude2}).

The above values of the bending modulus $\kappa$ are somewhat larger
than the values commonly reported in experiments in phospholipids:
$\kappa\sim10-20\,kT$ \cite{seifert_lipowsky}. We should, in
principle, correct our values of $\kappa$ by considering the reduction
in the effectively measured bending modulus due to long-wavelength
thermal undulations. The correction term depends logarithmically on the
size of the system \cite{safran,peliti_leibler}:
\begin{equation}
\frac{\Delta \kappa}{\kappa}=\frac{3kT}{4\pi\kappa}\ln\left(
\frac{L}{a}\right),
\end{equation}
where $a$ is some microscopic length. Setting $a\sim l=L/8$ (the mesh
size in our simulations), and using the values of $\kappa$ in
Eqs.(\ref{moduli28.75}) and (\ref{moduli29.375}), we find that this
correction amounts to about $1\%$ of the value of $\kappa$ and,
therefore, falls within the uncertainty in our estimates of the
bending modulus. 

The fact that the surface tension $\gamma$ is negative has an
interesting implication: It means that the size of the membrane cannot
grow indefinitely, but an upper bound exists
\begin{equation}
L_c\lesssim 2\pi\sqrt{\frac{\kappa}{|\gamma|}}
\end{equation}
above which there are small $q$ modes with $|\vec{q}\,|<q_c=2\pi/L_c$
that make the system unstable [see Eq.(\ref{helfhamiltonian5})]. One
can also understand the origin of this instability in ``real'' space,
rather in $q$ space: When $\gamma<0$ the elastic energy of the
membrane {\em decreases}\/ by increasing its area, and it is the
bending energy that stabilizes the system. Modes with larger
wavelength (smaller $|\vec{q}\,|$) require smaller curvatures to
increase the area of the membrane and, thus, cost less bending energy.
Membranes with linear size $L>L_c$ have long wavelength modes that
reduce the elastic area energy more than they increase the elastic
bending energy. For systems whose linear size is smaller than, but
close to, $L_c$ the amplitude of the small $q$ modes become large [see
Eq.(\ref{qamplitude})], and the approximation in
Eqs.(\ref{helfhamiltonian4}) and (\ref{helfhamiltonian5}) is no longer
valid. It is then necessary to include higher order terms in the
Hamiltonian (\ref{helfhamiltonian5}), and to consider their influence
on the spectrum of thermal undulations and on the stability of the
membrane.

\subsection{Pore Formation}
\label{poresection}

The projected area $\tilde{A}_p$ at which the surface tension vanishes
is called the saturated ({\em Schulman}\/) area
\cite{schulman,degennes_taupin}. One can evaluate the saturated area
of our model membranes using a linear approximation for the relation
between the surface tension and the excess area $\delta
A_p=A_p-\tilde{A}_p$ \cite{remark}
\begin{equation}
\gamma=K_A\left(\frac{\delta A_p}{\tilde{A}_p}\right). 
\label{compressibility}
\end{equation}
The coefficient $K_A$ appearing in the above equations is the area
compressibility modulus of the membrane. Using the values of $\gamma$
obtained from the simulations for $A_p=(28.75)^2\simeq 826.6$, and
$A_p=(29.375)^2\simeq 862.9$ [Eqs.(\ref{moduli28.75}) and
(\ref{moduli29.375})] in Eq.(\ref{compressibility}), we derive the
following estimates:
\begin{equation}
K_A=19.6\pm 6.6\ \frac{kT}{\sigma^2},
\label{numcompressibility}
\end{equation}
and, 
\begin{equation}
\tilde{A}_p=890\pm 17,
\label{saturated}
\end{equation}
for $K_A$ and $\tilde{A}_p$. Equation (\ref{saturated}) suggests that
our membrane with $A_p=(30.0)^2=900$ might be found above the
saturated area $\tilde{A}_p$ and, therefore, may have a positive
surface tension. The attempt to verify this conjecture by analyzing
the spectrum of the membranes, as done for the fluid membranes with
lower projected areas, is hampered by two technical difficulties. The
first one is related to the flip-flop motion of lipids between the two
leaflets. The trans-bilayer diffusion makes it computationally
complicated to determine which lipids are related to which layer of
the membrane and, therefore, it becomes difficult to calculate the
profile of the layers. The other difficulty results from the holes
which are created in the membrane. These pores may have an area larger
than $a_0$, the area of the grid cells. In such a case we find an
empty cell with no lipids inside, and the height of the membrane at
the corresponding grid point cannot the evaluated (unless one
interpolates this value using the height of the membrane at the
adjacent grid points). We have taken advantage of the fact that the
typical time for the appearance of the pores and for the flip-flop
motion which accompanies their formation, is larger than the
relaxation time of the spectrum, and used short MC runs (during which
pores were not observed) to estimate the surface tension of the
membrane. We found a positive surface tension with a magnitude of the
order of $\gamma\sim 1kT/\sigma^2$, which is roughly half an order of
magnitude larger than the value anticipated by
Eqs.(\ref{compressibility})-(\ref{saturated}). While the spectral
analysis supports our conjecture that $\gamma$ is positive for
$A_p=(30.0)^2$, one should not attempt to use many independent short
runs to achieve a more accurate estimate of $\gamma$. It is unclear
how well equilibrated the membranes in these short MC runs are.
Moreover, it is incorrect to base such an estimate on statistical
averaging restricted to membranes without pores. The creation of the
pores tends to reduce the surface tension since they make the
effective area of the membrane smaller.

Membranes with a positive surface tension can reduce their elastic
energy by decreasing their area, and the formation of pores is
obviously one of the mechanisms to achieve that. Other ways of
reducing the membrane area which are not possible in our model is to
decrease the projected area or to increase the area density by
adsorbing lipids from the solvent. For the case of a pore formation,
one has to consider the line tension energy price involved with the
creation of the hole. The simplest theoretical model discussing pore
formation was suggested by Litster \cite{litster}. In this
zero-temperature model, the contribution of a circular hole of radius
$R_{\rm pore}$ to the free energy of a membrane with a positive
surface tension $\gamma$ is given by
\begin{equation}
F_{\rm pore}=-\gamma \pi R_{\rm pore}^2 +\lambda 2\pi R_{\rm pore},
\label{litstermodel}
\end{equation}
where $\lambda$ is the line tension of the hole. According to this
model a pore with a radius larger than the critical value of
$\lambda/\gamma$ is predicted to grow without bound. Such a
thermodynamically large circular hole can be created only if the
critical energy barrier $\pi\lambda^2/\gamma$ is accessible by thermal
fluctuations. At a finite temperature it is necessary to take into
account the entropy of the pores and the picture becomes more
complicated. Recent computer simulations \cite{schillcock} have
demonstrated that the typical shape of thermally induced pores is
non-circular but rather of a self-avoiding ring or a branched polymer.
The most striking feature predicted by this study was, however, the
fact that pores can appear at zero, and even at small negative surface
tension.

The major drawback of the above model is the fact that while it
predicts the expansion of the pore without limit, the first term in
Eq.(\ref{litstermodel}), assuming a linear relation between the
reduction in elastic energy and the area of the pore, applies to small
pores only. An improved model can be obtained by assuming other forms
of the free energy dependence on the pore area. We first consider a
zero temperature model where the membrane does not fluctuate in the
normal direction. The free energy of the membrane (which at zero
temperature coincides with the potential energy) has a minimum at the
saturated area which we shall now denote by $\tilde{A}^E$ [compare
this notation with the one used in Eq.(\ref{compressibility})] to
indicate that it is determined by energy consideration. The subscript
$p$ has been omitted since the projected area is also the total area
of the membrane in this case. Close to $\tilde{A}^E$ we can use the
quadratic approximation to describe the dependence of the free energy
density $f$ on the excess area $\delta A^E=A-\tilde{A}^E$
\begin{equation}
f\equiv\frac{F}{\tilde{A}^E}=
\frac{1}{2}K_A^E\left(\frac{\delta A^E}{\tilde{A}^E}\right)^2,
\label{energetic}
\end{equation}
where, as in the case of the saturated area, we use the superscript
$E$ in the notation of the area compressibility $K_A^E$. If a pore of
area $A_{\rm pore}$ is formed then the area of the membrane is reduced
by $A_{\rm pore}$ and, consequently, the pore contribution to the free
energy density is given by:
\begin{eqnarray}
f_{\rm pore}\left(A,A_{\rm pore}\right)
&=&\frac{1}{2}K_A^E\left(\frac{\delta A^E-A_{\rm pore}}
{\tilde{A}^E}\right)^2
\nonumber \\
\ &-&
\frac{1}{2}K_A^E\left(\frac{\delta A^E}{\tilde{A}^E}\right)^2
\nonumber \\
\ &+&\frac{2\lambda\sqrt{\pi}}{\tilde{A}^E}\sqrt{A_{\rm pore}}.
\label{ourmodel}
\end{eqnarray}
As in Eq.(\ref{litstermodel}), we consider a circular hole and, thus,
its perimeter and area are related by $\Gamma=\sqrt{4\pi A}$. The
equilibrium size of the pore $A^*_{\rm pore}$ is found by solving the
equation $\partial f_{\rm pore}/\partial A_{\rm pore}=0$, and in
addition by verifying that $f_{\rm pore}\left(A,A^*_{\rm
    pore}\right)<f_{\rm pore}\left(A,0\right)=0$. While in Litster's
model a membrane with positive surface tension can be only metastable
against the formation of a pore, the model presented here yields a
different scenario: Pores are thermodynamically unfavorable as long as
the line tension satisfies
\begin{equation}
\lambda>\lambda'=
\sqrt{\frac{2}{27\pi}}\frac{({\delta A^E})^{3/2}}{\tilde A^E}K_A^E.
\label{critical}
\end{equation}
At this value a first order first transition occurs, and a pore of
size $A^*_{\rm pore}=2/3\,\delta A^E$ is created. The pore grows
gradually as $\lambda$ is decreased below this value. When
$\lambda\rightarrow 0$, $A^*_{\rm pore}\rightarrow \delta A^E$, and
the effective area of membrane attains the optimal (Schulman) value
$\tilde{A}^E$. As in Litster's model, there exists a free energy
barrier for the formation of the pore. At the transition
($\lambda=\lambda'$) the height of the barrier is
\begin{equation}
\Delta F \sim \lambda^{4/3}(\delta
A^E)^{1/3}/{K_A^E}^{1/3}.
\label{barrier}
\end{equation}

A theory for the entropic contribution to the free energy of the pore
has been recently presented by Sens and Safran \cite{sens_safran}.
According to this theory, hole formation is one of the mechanisms to
``redeem'' the degrees of freedom associated with the long wavelength
modes in the fluctuation spectrum which are eliminated by the surface
tension. To a first approximation, the effect of this entropic surface
tension can be easily incorporated into Eqs.(\ref{energetic}) and
(\ref{ourmodel}). Let us assume for a moment that $K_A^E=0$, and that
the projected area of the membrane $A_p$ is fixed. Because of the
thermal fluctuations, the total area of the membrane will be larger
than $A_p$. As has been explained in Ref.~\cite{sens_safran}, there
exists an optimal total area at which the membrane is tensionless
\begin{equation}
\tilde{A}^S\simeq A_p\left[1+\frac{kT}{8\pi\kappa}
\ln\left(\frac{A_p}{l^2}\right)\right],
\label{entschulman}
\end{equation}
where $\kappa$ is the bending modulus, and $l$ is some molecular
cut-off length. The superscript $S$ denotes the fact that the optimal
area discussed here is entropic in nature, and does not need to be
equal to $\tilde{A}^E$ in Eq.(\ref{energetic}). A membrane with
$A\neq\tilde{A}^S$ will experience a surface tension. The free energy
associated with this entropic surface tension can be calculated
analytically. Here, however, we shall use the quadratic approximation
in $\delta A^S\equiv A-\tilde{A}^S$
\begin{equation}
f\equiv\frac{F}{A_p}=\frac{1}{2}K_A^S\left(\frac{\delta A^S}{A_p}\right)^2,
\label{entropic}
\end{equation}
which is valid only close to the minimum of the free energy at
$\tilde{A}^S$. The entropic area compressibility in
Eq.(\ref{entropic}) is given by \cite{sens_safran}
\begin{equation}
K_A^S=\frac{32\pi^3\kappa^2}{A_pkT}.
\label{entcompressibility}
\end{equation}
Combining the energetic (\ref{energetic}) and the entropic
(\ref{entropic}) contributions to the free energy, we find another
quadratic form for the {\em total}\/ free energy of the membrane
\begin{equation}
f\left(A\right)\equiv\frac{F}{A_p}=\frac{1}{2}K_A
\left(\frac{\delta A}{A_p}\right)^2,
\label{quadraticf2}
\end{equation}
where the excess area $\delta A=A-\tilde{A}$ is defined with respect
to the minimum at
\begin{equation}
\tilde{A}=\frac{K_A^E A_p+K_A^S\tilde{A}^S}{\frac{A_p}
{\tilde{A}^E}K_A^E+K_A^S},
\label{optimalarea}
\end{equation}
and the effective area compressibility is equal to
\begin{equation}
K_A=K_A^E\frac{A_p}{\tilde{A}^E}+K_A^S.
\label{effcompressibility}
\end{equation}
The optimal area and the area compressibility appearing in the above
two equations (and which include both energetic and entropic
contributions) should replace their purely energetic counterparts in
Eq.(\ref{ourmodel}) for the pore free energy density and in
Eqs.(\ref{critical}) and (\ref{barrier}) for the critical line tension
and the free energy barrier. For typical values of phospholipids:
$\kappa=10kT\sim 5\times10^{-13}$ ergs, and $A_p=(10\ \mu
m)^2=10^{-6}\ {\rm cm^2}$, we get upon substitution in
Eq.(\ref{entcompressibility}), $K_A^S\sim 5\times10^{-3}\ {\rm
  ergs/cm^2}$. This value of $K_A^S$ is several orders of magnitude
smaller than the area compressibility typically found in stretching
experiments $K_A\gtrsim 10^2\ {\rm ergs/cm^2}$ \cite{seifert_lipowsky}
and, therefore, the entropic contribution to $K_A$ and $\tilde{A}$ can
be neglected \cite{remark2}. The effect of the thermal fluctuations
can be felt only in small membranes (such as in this paper), as
$K_A^S$ increases by decreasing the projected area.  Unfortunately,
the membranes in our computer study are {\em too small}\/ to allow the
description of this effect by
Eqs.(\ref{entschulman})--(\ref{entcompressibility}). The derivation of
these equations is based on the assumption that the long wavelength
behavior of the membrane is dominated by the surface tension. This
requires that $\kappa(2\pi/\sqrt{A_p})^2\ll\gamma$ - a criterion
which is not satisfied in our case. The long wavelength fluctuations
in our membranes are mainly controlled by the curvature elasticity.

Is the appearance of pores in the simulations in accord with the model
described by Eq.(\ref{ourmodel})? In order to answer this question we
need to evaluate the line tension $\lambda$ of the pore. The line
tension $\lambda$ has the dimensions of energy per unit length. Its
magnitude can be estimated by noting that the lipids on the rim of the
pore have 1-2 less neighbors compared to the other lipids. Therefore,
the energy cost associated with each such lipid is of the order of the
interaction energy between two adjacent molecules which is roughly
$kT$. The length occupied by each lipid along the perimeter of the
hole is of the order of $\sigma$, and so $\lambda\sim kT/\sigma$. This
value of $\lambda$ should be smaller than the critical value
$\lambda'$ given by Eq.(\ref{critical}). Using the values of $K_A$ and
$\tilde{A}$ provided by Eqs.(\ref{numcompressibility}) and
(\ref{saturated}), we arrive at the estimate $\lambda'\sim kT/\sigma$
for the membrane with $A_p=30^2=900$ ($\delta A\sim 10$). This means
that $\lambda$ and $\lambda'$ are of the same order of magnitude and,
thus, may obey the criterion given by Eq.(\ref{critical}) for the
thermodynamic stability of membranes with holes. The fact the pores in
our simulations appear for only short time intervals, before they
close up, may indicate that $\lambda$ is, in fact, slightly larger
than $\lambda'$, and that the pores are only metastable. In addition
to to the values of $\lambda$ and $\lambda'$, we also need to check
the free energy barrier for the formation of the pores, as given by
Eq.(\ref{barrier}). We find $\Delta F\sim kT$, and so the opening of a
pore can be nucleated by thermal fluctuations.

\section{Summary and Discussion}
\label{summary}

We have introduced a new simple computer model for bilayer membranes
whose main feature is the fact that the system is simulated in vacuum
rather than in aqueous environment. The elimination of the solvent
from the simulations greatly improves computational efficiency.
Devising a ``water-free'' model is a great challenge as the water
molecules, via the electrostatic interactions between them and the
lipids, play a central role in the aggregation and the stabilization
of the membrane through the resulting hydrophobic effects.  The
self-assembly of the system has not been investigated in this paper.
(The reader is referred to the simulations presented in
Refs.~\cite{drouffe} and \cite{noguchi} in which this issue has been
addressed). We did, however, demonstrated that bilayers, once they are
formed, can be stable without the surrounding solvent. One only needs
to modify the interactions between the lipids, and use effective
potentials that compensate for the absence of water by producing a
barrier against the disintegration of the membrane. In this model we
have been able to stabilize the membrane using pair-wise short range
interactions only - another feature that reduces the computational
effort involved with the simulations. To the best of our knowledge,
this is the first water-free computer model in which fluid membranes
are being observed without the need of multibody interactions.

We have found that our simple model reproduces many known features of
bilayer membranes, such as the transition from a high density solid
phase to a low density fluid phase. We have inspected the spectrum of
thermal fluctuations of the fluid membranes, and found it to be well
described by the Helfrich Hamiltonian. From the analysis of the
spectral intensity of the different modes, we have extracted the
surface tension and bending modulus of the system. Based on our
numerical results for the surface tension, we have attempted to
determine the optimal area of the membrane at which the surface
tension vanishes.  Indeed, for areas larger than the saturated area,
we have found evidences that the surface tension becomes positive.
Fluid membranes with positive surface tension can develop pores, and
the creation of pores allows the diffusion of lipids from one layer to
the other (flip-flops). The opening of holes in our membranes is in
agreement with a simple model that takes into account the
contributions to the area compressibility of both the inter-particle
forces and the thermal fluctuations.

In order to make a closer contact with biological systems, it is
necessary to extend the model presented here to include the other
elements found in biomemebranes such as the membranes proteins and the
cytoskeleton. It would be interesting to see whether these additional
components can also be modeled in a coarse-grain manner that would
minimize both the computational and the conceptual complexity. Such a
model may shed light on an abundance of challenging problems like the
effect of the cytoskeleton on the elastic properties of the bilayer, or
the role played by the membrane proteins in transport processes across
the membrane.

We thank David Andelman, Grace Brannigan, Daan Frenkel, and Thomas
Powers for useful discussions, and to Niels Gr{\o}nbech Jensen and Kurt Kremer
for their comments on the manuscript. Special thanks to Philip Pincus
and Claus Jeppesen for their substantial contribution to the work, and
to Jeffrey Barteet for technical support. This work was supported by
the National Science Foundation under Award No.~DMR-0203755. The
Materials Research Laboratory at UC Santa Barbara is supported by NSF
No.~DMR-0080034.



\begin{thebibliography}{99}
\bibitem{remark0} Note that we use the definition of the diffusion 
constant in two-dimensional systems ($D\equiv \Delta r'(t)^2/4t$), 
rather than it's three-dimensional counterpart ($D\equiv \Delta 
r'(t)^2/6t$). The reason for this choice of a definition is the lack of 
diffusion in the direction normal to the membrane plane (as discussed in 
the paragraph).
\bibitem{remark2} For $A<\tilde{A}^S$ the entropic contribution to the
  free energy grows exponentially with $\delta A^S$ (see
  Ref.~{\protect \cite{sens_safran}}) and, thus, far from
  $\tilde{A}^S$ it may not be ignored.
\end{thebibliography}
\end{document}